# Multi-Proxy Multi-Signcryption Scheme from Pairings

LIU Jun-Bao, XIAO Guo-Zhen, *Member, IEEE*

*Abstract*—A first multi-proxy multi-signcryption scheme from pairings, which efficiently combines a multi-proxy multi-signature scheme with a signcryption, is proposed. Its security is analyzed in detail. In our scheme, a proxy signcrypter group could be authorized as a proxy agent by the cooperation of all members in the original signcrypter group. Then the proxy signcryptions can be generated by the cooperation of all the signcrypters in the authorized proxy signcrypter group on behalf of the original signcrypter group. The correctness and the security of this scheme are proved.

## I. INTRODUCTION

In 1996, Mambo et al [1] first introduced the concept of a proxy signature scheme, which allows the original signer to delegate his signature power to a designed signer, called the proxy signer. Then the proxy signer is able to sign on behalf of the original signer. Since the proxy signature scheme was brought forward, lots of schemes have been proposed [2-8]. In 2002, Hwang et al [3] first proposed the concept of multi-proxy multi-signature. In a multi-proxy multi-signature scheme, only the cooperation of all members in the original signer group could authorize a proxy group as his proxy agent. Then only the cooperation of all the signers in the authorized proxy group can generate the proxy signatures on behalf of the original signer group.

Signcryption is a public key primitive proposed by Zheng [9] to achieve the combined functionality of digital signature schemes and encryption in an efficient manner. LI et al [10] have proposed a proxy signcryption which efficiently combines a proxy signature scheme with a signcryption. However, none of the existing signcryption schemes is multi-proxy multi-signcryption scheme. In this paper, based on the Xue et al Scheme [8], we propose a multi-proxy multi-signcryption scheme from pairings. In a multi-proxy multi-signcryption scheme, only the cooperation of all members in the original signcrypter group could authorize a proxy group as his proxy agent. Then only the cooperation of all the signcrypters in the authorized proxy group can generate the proxy signcryptions on behalf of the original signcrypter group. We analyze the proposed scheme from correctness and security points of view. We show that the proposed scheme provide all the security properties required by multi-proxy multi-signcryption.

This paper is organized as follows. In Section 2, we define the notation of the bilinear pairing as preliminary. In Section 3, we propose our scheme. In Section 4, we discuss our scheme in terms of security. Finally, we conclude our paper in Section 5.

## II. PRELIMINARIES

In this section, we will briefly describe the basic definition and properties of the bilinear pairings and some problems. Let $G_1$ be a cyclic additive group generated by P, whose order is a prime q, and $G_2$ be a cyclic multiplicative group of the same order q. $H(\cdot)$ denotes a cryptography hash function and $x \in_R M$ means x is selected randomly in M. A bilinear pairings is a map $\hat{e}: G_1 \times G_1 \to G_2$ with the following properties:

1. Bilinear: For all $P, Q, R \in G_1$ and $a, b \in Z_q^*$, such that

$$\hat{e}(P, Q+R) = \hat{e}(P,Q)\hat{e}(P,R)$$

$$\hat{e}(P+Q, R) = \hat{e}(P,R)\hat{e}(Q,R)$$

$$\hat{e}(aP, bQ) = \hat{e}(abP, Q) = \hat{e}(P, abQ) = \hat{e}(P, Q)^{ab}$$

2. Non-degenerate: There exists $P \in G_1$, such that $\hat{e}(P, P) \neq 1$.

3. Computable: Given $P, Q \in G_1$, there is an efficient algorithm to compute $\hat{e}(P, Q)$.

Now we describe some mathematical problems in relation to bilinear pairs.

Discrete Logarithm Problem ($DLP$): Given two group elements $P$ and $Q$, find an integer $n$, such that $Q = nP$ whenever such an integer exists.

Decision Diffie-Hellman Problem ($DDHP$): For $a, b, c \in_R Z_q^*$, $P \in G_1$, given $P, aP, bP, cP$, decide whether $c = ab \mod q$.

Computational Diffie-Hellman Problem ($CDHP$): For $a, b \in_R Z_q^*$, $P \in G_1$, given $P, aP, bP$, compute $abP$.

Gap Diffie-Hellman Problem ($GDHP$): On the group $G_1$, $DDHP$ is easy but $CDHP$ is hard. Now we call $G_1$ a Gap Diffie-Hellman ($GDH$) group.

We assume through this paper that $CDHP$ and $DLP$ are intractable in $G_1$ and $G_2$. $GDH$ groups can be found on supersingular elliptic curves or hyperelliptic curves over finite field, and the bilinear parings can be derived from the Weil or Tate pairing. Our scheme is based on $GDH$ group.

Basically, a secure multi-proxy multi-signcryption scheme should satisfy the following requirements: Strong unforgeability, Verifiability, Strong identifiability, Strong undeniability, Prevention of misuse, and Confidentiality.

### III. PROPOSED SCHEME

This section proposes a pairing-based multi-proxy multi-signcryption scheme. Our multi-proxy multi-signcryption scheme is divided into four phases: System initialization phase, Proxy key generation phase, multi-proxy multi-signcryption generation phase, and Unsigncryption phase.

*System Initialization*: Let $A_1 \cdots A_n$ be $n$ original signcrypters. For $1 \leq \forall i \leq n$, $A_i$ has a private key $x_{ai}$ and corresponding public key $y_{ai}$, such that $x_{ai} \in_R Z_q^*$ and $y_{ai} = x_{ai}P$. Let the proxy group consist of $l$ proxy signcrypters. The proxy signcrypter $P_j (j = 1, 2, \ldots, l)$ owns its private key $x_{pj}$ and its public key $y_{pj}$, such that $x_{pj} \in_R Z_q^*$ and $y_{pj} = x_{pj}P$. The original signcrypters $A_i (i = 1, 2, \ldots, n)$ jointly ask the proxy signcrypter group $P_j (j = 1, 2, \ldots, l)$ to carry out signcrypting a document $m$ for them altogether. Let $C$ be the unsigncrypter. $C$ has a secret key $x_c$ and corresponding public $y_c$, such that $x_c \in_R Z_q^*$ and $y_c = x_c P$. Let E and D be the encryption and the decryption functions, respectively, defined by an available symmetric algorithm, and be previously known to the signcrypter, the proxy signcrypter and the unsigncrypter. Let public collision resistant hash function $H_1(\cdot)$, $H_2(\cdot)$ and $H_3(\cdot)$ be $H_1 : \{0,1\}^* \to Z_q^*$, $H_2 : \{0,1\}^* \to G_1$, $H_3 : G_2 \to \{0,1\}^n$ respectively. Finally the system parameters are $(G_1, G_2, \hat{e}, P, q, H_1, H_2, H_3, E, D)$.

*Proxy Key Generation*: To delegate the signcrypting capability to a group of proxy signcrypters, the original signcrypters do the following to make the signed warrant $m_w$. In $m_w$, there is an explicit description of the delegation relation including the identity of the original signcrypters and the proxy signcrypters, the message to be signed, and so on. If the following process is finished successfully, each proxy signcrypter $P_j (j = 1, 2, \ldots, l)$ gets a proxy key $S_{pj} (j = 1, 2, \ldots, l)$.

1. Each original signcrypter computes $S_{ai} = x_{ai} H_2(m_w)$ and broadcasts his $S_{ai}$ to the $l$ proxy signcrypter. In addition, the first original signcrypter broadcasts $m_w$ to the $l$ proxy signcrypter, for

$i = 1,2,\ldots,n$.

2. The proxy signcrypter group verifies the correctness of $S_{ai}$ by an equation

$$\hat{e}(P, S_{ai}) = \hat{e}(y_{ai}, H_2(m_w)) \quad (1)$$

for $i = 1,2,\ldots,n$.

3. Once all of the above Eq. (1) hold, the proxy signcrypter group computes $S_A = \sum_{i=1}^{n} S_{ai}$. Each member of the proxy signcrypter group $P_j$ computes his proxy signcryption key

$$S_{pj} = S_A + x_{pj} H_2(m_w) \quad (2)$$

for $j = 1,2,\ldots,l$.

*Multi-Proxy Multi-Signcryption Generation*: To generate a multi-proxy multi-signcryption on a message $m$ that conforms to the warrant $m_w$, one proxy signcrypter in the proxy signcryption group is designated as a clerk, whose task is to combine partial proxy signcryption to generate the final multi-proxy multi-signcryption.

1. Each proxy signcrypter $P_j$ selects an integer $t_j \in_R Z_q^*$, for $j = 1,2,\ldots,l$.

2. Each proxy signcrypter $P_j$ computes $r_{pj} = \hat{e}(P, y_c)^{t_j}$ and broadcasts $r_{pj}$ to the other $l-1$ proxy signcrypters, for $j = 1,2,\ldots,l$.

3. Each proxy signcrypter $P_j$ computes $k = H_3(\prod_{j=1}^{l} r_{pj})$, $c = E_k(m)$, $r_p = H_1(c \| k)$, $u_{pj} = t_j P - r_p S_{pj}$, and sends $u_{pj}$ to the clerk as his partial proxy signcryption on $m$, for $j = 1,2,\ldots,l$.

4. The clerk computes $u_p = \sum_{j=1}^{l} u_{pj}$, $S = lS_A$, and sends $(m_w, S, c, r_p, u_p)$ to the unsigncrypter.

*Unsigncryption*: After receiving the multi-proxy multi-signcryption $(m_w, S, c, r_p, u_p)$, $C$ computes

$$k = H_3(\hat{e}(u_p, y_c)\hat{e}(S, y_c)^{r_p} \hat{e}(H_2(m_w), \sum_{j=1}^{l} y_{pj})^{r_p x_c}) \quad (3)$$

$$m = D_k(c) \quad (4)$$

and accepts the multi-proxy multi-signcryption if and only if $r_p = H_1(c \| k)$.

IV. SECURITY ANALYSIS

*Verifiability*: The consistency of this scheme can be verified as follows:

$$\begin{aligned} k &= H_3(\prod_{j=1}^{l} r_{pj}) \\ &= H_3(\prod_{j=1}^{l} \hat{e}(P, y_c)^{t_j}) \\ &= H_3(\hat{e}(\sum_{j=1}^{l}(u_{pj} + r_p S_{pj}), y_c)) \\ &= H_3(\hat{e}(u_p, y_c)\hat{e}(\sum_{j=1}^{l}(S_A + x_{pj} H_2(m_w)), y_c)^{r_p}) \\ &= H_3(\hat{e}(u_p, y_c)\hat{e}(S, y_c)^{r_p} \hat{e}(H_2(m_w), \sum_{j=1}^{l} y_{pj})^{r_p x_c}) \end{aligned} \quad (5)$$

In the unsigncryption phase, $C$ can recover the value of $k$, thus the signcryption can be unsigncrypt.

*Strong Unforgeability*: As for multi-proxy multi-signcryption, there are mainly four kinds of attackers: any third party, who do not participate the issue of the multi-proxy multi-signcryption; some proxy signcrypter, who play an active in the signcrypting process; the original signcrypter and the signcryption owner. Because the multi-proxy multi-signcryption $u_p = \sum_{j=1}^{l} u_{pj}$ contains secret key information $x_{pj}$ of each proxy signcrypter $P_j$ in the proxy multi-signcryption key generation phase, without secret key information $x_{pj}$ of $P_j$, any third party, some proxy signcrypter, the signcryption owner and the original signcrypters cannot generate a valid multi-proxy multi-signcryption scheme by themselves.

*Strong Identifiability*: The unsigncrypter can distinguish proxy's normal signcryption from his multi-proxy multi-signcryption, because the multi-proxy multi-signcryption key is different from his own private key.

*Strong Nonrepudiation*: In our scheme, each proxy signcrypter

$P_j$ cannot repudiate his participation on multi-proxy multi-signcryption while illegal attacker cannot claim that he is proxy signcrypter, because $u_p = \sum_{j=1}^{l} u_{pj}$ contains secret key information $x_{pj}$ of each proxy signcrypter $P_j$, at the same time, warrant $m_w$ also contains identity information of $P_j$, In addition $k = H_3(\hat{e}(u_p, y_c)\hat{e}(S, y_c)^{r_p}\hat{e}(H_2(m_w), \sum_{j=1}^{l} y_{pj})^{r_p x_c})$ contains $y_{pj}$ and $m_w$.

*Confidentiality*: Because the secret key $k$ contains secret key information $x_c$ of $C$, only $C$ can compute $k$ and recover m.

*Prevention of Misuse*: Each proxy signcrypter $P_j$ cannot repudiate his participation on multi-proxy multi-signcryption, in addition, $m_w$ includs the message to be signed, so our scheme can prevent the misuse.

## V. CONCLUSIONS

In this paper, based on the Xue et al Scheme [8], we construct a new multi-proxy multi-signcryption scheme from pairings (To the best of our knowledge, there is not any multi-proxy multi-signcryption scheme published in the open literature.) and analyze its security in detail. We prove that this scheme is correct and secure.

## ACKNOWLEDGMENT

The authors would like to thank the anonymous reviewers for their helpful comments.